\documentclass{LMCS}

\def\dOi{9(3:9)2013}
\lmcsheading%
{\dOi}
{1--11}
{}
{}
{Jan.~14, 2013}
{Sep.~\phantom03, 2013}
{}

\usepackage{amsmath,amssymb,enumerate,hyperref}

\theoremstyle{plain}\newtheorem{example}[thm]{Example}

\def\MC{{\mathbb C}}
\def\ME{{\mathbb E}}
\def\MP{{\mathbb P}}
\def\MR{{\mathbb R}}
\def\CB{{\mathcal B}}

\begin{document}

\title[Computability of Probability Distributions
and Characteristic Functions]
{Computability of Probability Distributions \\ 
and Characteristic Functions}

\author[T.~Mori]{Takakazu Mori\rsuper a} 
\address{{\lsuper{a,b}}Faculty of Science, Kyoto Sangyo University}
\email{\{morita,tsujiiy\}@cc.kyoto-su.ac.jp}
\thanks{{\lsuper a}This work has been supported in part by Scientific Foundations of 
JSPS No. 21540152.}  

\author[Y.~Tsujii]{Yoshiki Tsujii\rsuper b} 
\address{\vskip-6 pt}
%\email{tsujiiy@cc.kyoto-su.ac.jp}
\thanks{{\lsuper b}This work has been supported in part by Scientific Foundations of 
JSPS No. 23540170.}  

\author[M.~Yasugi]{Mariko Yasugi\rsuper c}
\address{{\lsuper c}Kyoto Sangyo University and Kyoto University}
\email{yasugi@cc.kyoto-su.ac.jp}


\ACMCCS{[{\bf Theory of computation}]: Models of
  computation---Computability---Recursive functions}

\begin{abstract}
As a part of our works on effective properties of probability 
distributions, we deal with the corresponding characteristic 
functions.  
A sequence of probability distributions is computable if and only if 
the corresponding sequence of characteristic functions is computable. 
As for the onvergence problem, the   effectivized Glivenko's theorem holds.  
Effectivizations of Bochner's theorem and de Moivre-Laplace 
central limit theorem are also proved.
\end{abstract}

\maketitle

\section{Introduction}

We are concerned with mutual relationships 
between computability of probability distributions 
(Borel probability measures on the real line $\MR$)
and that of the corresponding probability distribution functions 
as well as that of the corresponding characteristic 
functions.  We are also concerned with mutual relationship between 
effective convergence of probability distributions 
and that of the corresponding probability distribution functions 
as well as that of the corresponding characteristic 
functions.  

We have no general characterization of  probability distribution 
functions which correspond to computable probability distributions.  
What we can do is to start with computable 
(hence continuous) probability distribution functions (\cite{MTY09b}) 
and examine what will happen with wider classes of 
probability distribution functions. 
In \cite{MTY09b}, it has been proved that, for a sequence of probability 
distributions with effectively bounded densities, 
it is computable if and only if the 
corresponding sequence of probability distribution functions is 
computable, and that it converges effectively to a (computable) 
probability distribution if and only if the corresponding sequence of probability 
distribution functions behaves similarly.

However, the Dirac probability distribution has no density function.  
Furthermore, its probability distribution function is discontinuous. 
We have then treated a wider class of probability distribution functions, 
namely Fine computable ones (cf. \cite{MTY10}, \cite{MTY11}), 
as the first step to deal with cases where probability 
distributions may not have bounded density functions.  

Here in this article, we will work on the relationship between 
computability of probability distributions and that of the corresponding 
characteristic functions as well as between their effective 
convergences.  Effective versions of some classical results 
regarding probability distributions and the corresponding characteristic 
functions such as Glivenko's theorem and de Moivre-Laplace 
central limit theorem will be proved. 

Let us explain some facts on probability distributions and 
related objects.

Many theorems in probability theory concerning convergence of random variables, 
such as {\it central limit theorems},  
are formulated in terms of convergence of probability distributions, 
and convergence of probability distributions is proved in terms of convergence 
of the corresponding characteristic functions. 

For a probability distribution $\mu$, its probability distribution function 
is defined by \\ 
$F(x)=\mu((-\infty,x])$, and characterized by the following properties: 
\begin{enumerate}[(Fi)]
\item monotonically increasing, 
 
\item right continuous, 

\item $\lim_{x\to-\infty}F(x)=0$ and $\lim_{x\to\infty}F(x)=1$.

\end{enumerate}

\noindent On the other hand, the characteristic function of $\mu$ is defined 
as the Fourier transformation, that is, 
$\varphi(t)=\int_{\MR}e^{itx}\mu(dx)=\mu(e^{it\cdot})$, 
and characterized by the following:  
\begin{enumerate}[(Ci)]

\item $\varphi$ is positive definite. 

\item $\varphi(t)$ is continuous at 0.  

\item $\varphi(0)=1$. 
\end{enumerate}

\noindent To obtain a probability distribution from a probability distribution 
function amounts to a construction of a measure.  To obtain 
a probability distribution from a characteristic function corresponds 
to Bochner's Theorem.

In this article, we investigate relations between computability of 
probability distributions and computability of the corresponding 
characteristic functions as well as relations between effective convergence 
of probability distributions and effective convergence of the 
corresponding characteristic functions.

In Section 2, we review briefly the previous results in \cite{MTY09b} and 
\cite{MTY11}.   

In Section 3, we treat effective integrability on $\MR$ and 
generalize some theorems 
in \cite{MTY08a}, \cite{MTY08b} and \cite{MTY10} to turn them to 
account for our purpose.  

Section 4 contains most of our main results.  We will  work on some  
relations between effective 
properties of a sequence of probability distributions $\{\mu_m\}$ and the 
corresponding sequence of characteristic functions $\{\varphi_m\}$.  
We have the following results:  
equivalence of computability of 
$\{\mu_m\}$  and that of the corresponding 
$\{\varphi_m\}$ (Theorems 4.1 and 4.5); 
effective Glivenko's theorem on convergence (Theorems 4.2 and 4.3) 
and  effective Bochner's theorem (Theorem 4.7). 

In section 5, we apply Theorem \ref{ThEffGlivenko} to prove the effective 
 de Moivre-Laplace central limit theorem.

\section{Preliminaries}

We follow the notions of computability developed by Pour-El and 
Richards (\cite{PR88}). Computability of a sequence of functions and 
effective convergence of a sequence of functions are formulated as follows. 

\begin{defi} \ ({\em Computability of a sequence of functions 
on $\MR$  \cite{PR88}}) \ 
A sequence of functions $\{f_n\}$ on $\MR$ is said to be computable if 
it satisfies the following two properties. 
\begin{enumerate}[(i)]
\item Sequential computability: the double sequence $\{f_n(x_m)\}$ is 
computable for any computable sequence $\{x_m\}$.  

\item Effective continuity: there exists a recursive function 
$\alpha(n,L,k)$ such that 
$|x-y|<2^{-\alpha(n,L,k)}$ and $|x|,|y|\leq L$ 
imply $|f_n(x)-f_n(y)|<2^{-k}$. 
\end{enumerate}
\end{defi}

\begin{defi} \ 
({\em Effective convergence of a sequence of functions on $\MR$ \cite{PR88}})\ 
A double sequence of functions $\{f_{n,m}\}$ converges effectively to 
a sequence $\{f_n\}$ if there exists a recursive function $\alpha(n,L,k)$ 
such that 
$m\geq \alpha(n,L,k)$ and $|x|\leq L$ imply $|f_{n,m}(x)-f_n(x)|<2^{-k}$. 
\end{defi}

We say that a sequence of functions $\{f_n\}$ is computable 
with compact support $L(n)$, if $\{f_n\}$ is (uniformly) computable 
and $L(n)$ is a recursive function which satisfies that 
$f_n(x)=0$ for all $x$ with $|x|\geq L(n)$. 

A sequence of functions $\{f_n\}$ is said to be effectively bounded 
if there exists a recursive function $M(n)$ which satisfies 
$|f_n(x)|\leq M(n)$ for all $x$. 

We denote $\int_{\MR}f(x)\mu(dx)$ with $\mu(f)$. 

\begin{defi}\label{DfComSqPD}\ ({\em Computability of 
probability distributions  \cite{MTY09b}}) \ % ,\cite{MTY11}
A sequence of probability distributions $\{\mu_m\}$ is  said to be computable, 
if $\{\mu_m(f_n)\}$ is computable for any computable $\{f_n\}$ 
with compact support $L(n)$. 
\end{defi}

\begin{defi}\label{DfConvPD}\ ({\em Effective convergence of 
probability distributions  \cite{MTY09b}}) \ % ,\cite{MTY11}
A sequence of probability distributions $\{\mu_m\}$ 
is said to converge effectively to a probability distribution 
$\mu$ if, for any computable sequence of functions $\{f_n\}$ with compact support, 
there exists a recursive function $\alpha(n,k)$ such that 
\begin{equation}\label{EqEffConv}
|\mu_m(f_n)-\mu(f_n)|<2^{-k} \mbox{\ if \ } m\geq\alpha(n,k)
\end{equation}  
holds. 
\end{defi}

If the condition (\ref{EqEffConv}) holds, 
we say that $\{\mu_m(f_n)\}$ converges effectively to 
$\mu(f_n)$ as $m$ tends to infinity effectively in $n$. 

The following proposition holds.

\begin{prop}\label{PpLimEffComvDist} \ {\rm(\cite{MTY09b})} \ 
If a computable sequence of probability distributions $\{\mu_m\}$ 
converges effectively to a probability distribution $\mu$, 
then $\mu$ is computable. 
\end{prop}

\begin{lem}\label{LmMonConv}\ {\em(Monotone Lemma  \cite{PR88})} \ 
Let $\{x_{n,k}\}$ be a computable sequence of reals which 
converges monotonically to $\{x_n\}$ as $k$ tends to infinity for each $n$. 
Then $\{x_n\}$ is computable if and only if the convergence is 
effective in $n$. 
\end{lem}

We will use the following functions $w_n$ and $w^c_n$. 

\vspace{1em} 

\begin{minipage}{21em}
$w_n(x):=\left\{\begin{array}{ll}
0 & \mbox{if \ } x \leq -n-1 \\
(x+n)+1  & -n-1 \leq x\leq -n \\ 
1         & \mbox{if \ }-n \leq x\leq n \\ 
-(x-n)+1 & \mbox{if \ }n\leq x\leq n+1 \\
0 & \mbox{if \ }x\geq n+1
\end{array}\right.. $

\vspace{0.8em}

$w^c_n(x):=1-w_n(x).$ 
\end{minipage}
\hspace*{1.3em}
\begin{minipage}{14em}
\begin{center}
\setlength{\unitlength}{0.4mm}
\begin{picture}(120,50)
\put( 5, 5){\line(1,0){130}}
\put( 5,35){\line(1,0){130}}
\put( 5, 5){\thicklines\line(1,0){10}}
\put(15, 5){\thicklines\line(2,3){20}}
\put(35,35){\thicklines\line(1,0){60}}
\put(95,35){\thicklines\line(2,-3){20}}
\put(115, 5){\thicklines\line(1,0){20}}
\put(35, 5){\line(0,1){30}}
\put(95, 5){\line(0,1){30}}

\put( -2,33){\footnotesize 1}
\put( -2, 3){\footnotesize 0}
\put(1, -3){\footnotesize$-n-1$}
\put(34, -3){\footnotesize$-n$}
\put(92, -3){\footnotesize$n$}
\put(105, -3){\footnotesize$n+1$}
\put(65,18){$w_n$}
\end{picture}

\vspace{1em}
Figure 1: $w_n(x)$ 
\end{center}
\end{minipage}

\vspace{1.5em}

\noindent$w_n$ and $w^c_n$ satisfy the following properties. 
\begin{enumerate}[$\bullet$]
\item $w_n(x)+w^c_n(x)\equiv 1$.

\item $\{w_n\}$ is monotonically increasing in $n$ 
and $\{w^c_n\}$ is monotonically decreasing. 

\item $\chi_{[-n,n]}(x)\leq w_n(x)\leq \chi_{[-(n+1),n+1]}(x)$. 

\item $\chi_{(-\infty,-(n+1)]}(x)+\chi_{[n+1,\infty)}(x)\leq w^c_n(x) \leq 
\chi_{(-\infty,-n]}(x)+\chi_{[n,\infty)}(x)$. 
\end{enumerate}

\begin{lem}\label{LmEffTight1}\ 
If a sequence of probability distributions $\{\mu_m\}$ is computable,
then there exists a recursive function $L(m,k)$ such that 
$\mu_m(w_n)> 1-2^{-k}$, or equivalently $\mu_m(w^c_n)<2^{-k}$, 
for all $n\geq L(m,k)$. 
\end{lem}

\noindent Observe that for all $n\ge L(m,k)$,
\[\mu_m([-(n+1),n+1])> 1-2^{-k}\quad\hbox{and}\quad 
  \mu_m((-\infty,-(n+1)])+\mu_m([n+1,\infty))<2^{-k}\ .
\]
  However, these quantities may not be computable.

\vspace{0.5em}

\noindent
{\it Proof of Lemma \ref{LmEffTight1}.} \  
By the definition, $\{\mu_m(w_n)\}$ is a computable sequence of reals and 
converges monotonically to $\mu_m(\MR)$ ($=1$) from below  as $n$ tends to infinity.
By Lemma \ref{LmMonConv}, this convergence is effective. 
\qed

\begin{lem}\label{LmEffTight}\ 
{\rm (Effective tightness of an effectively convergent sequence, 
Effectivization of Lemma 15.4 in \cite{Tsurumi64})} \ 
If a computable sequence of probability distributions $\{\mu_m\}$ 
effectively converges to a probability distribution $\mu$, then 
there exists a recursive function $\alpha(k)$ such that 
$\mu(w_{\alpha(k)}^c)<2^{-k}$ and 
$\mu_m(w_{\alpha(k)}^c)<2^{-k}$ for all $m$. 
It also holds that 
\[\mu_m([-\alpha(k)-1,\alpha(k)+1]^c)<2^{-k}\quad\hbox{for all $m$ and}\quad
  \mu([-\alpha(k)-1,\alpha(k)+1]^c)<2^{-k}\ .
\] 
\end{lem}

\noindent
{\it Proof.} \ 
Let ${L}(m,k)$ be the  recursive function in Lemma \ref{LmEffTight1}, 
and let ${L}(k)$ be the one for a single $\mu$. 

By definition, $\{w_n\}$ is a uniformly computable sequence of functions 
with compact support. 
So, $\{\mu_m(w_n)\}$ converges effectively to $\{\mu(w_n)\}$, 
that is, there exists a recursive function $\gamma(n,k)$ such that 
$m\geq\gamma(n,k)$ implies $|\mu_m(w_n)-\mu(w_n)|<2^{-k}$.  

It holds that $\mu(w^c_n)< 2^{-k}$ 
for all $n \geq {L}(k)$ by Lemma \ref{LmEffTight1}. 

If $m\geq \gamma({L}(k),k)$, then 
$|\mu_m(w^c_{{L}(k)})-\mu(w^c_{{L}(k)})|=|\mu_m(w_{{L}(k)})-\mu(w_{{L}(k)})|
<2^{-k}$ and hence $\mu_m(w^c_{{L}(k)})< 2\cdot 2^{-k}$. 
For each $\ell\leq\gamma({L}(k),k)-1$, 
$\mu_{\ell}(w^c_{{L}(\ell,k)})<2^{-k}$ by Lemma \ref{LmEffTight1}. \\ 
If we put 

$\alpha(k)=\max\{{L}(k+1), {L}(1,k+1), {L}(2,k+1),\ldots,
{L}(\gamma({L}(k+1),k+1)-1,k+1)\}$, \\ 
then $\mu(w_{\alpha(k)}^c)<2^{-k}$ and 
$\mu_m(w^c_{\alpha(k)})< 2^{-k}$ holds for all $m$. 
\qed

\begin{prop}\label{PpEqvWcomp2CUCompFt} \ 
{\rm(\cite{MTY09b})} \ % , \cite{MTY11}
If $\{\mu_m\}$ is computable, then it is {\rm weakly sequentially computable}, 
that is,  $\{\mu_m(f_n)\}$ is a computable sequence of reals 
for all effectively bounded computable sequence of functions $\{f_n\}$. 
\end{prop}

\begin{prop}\label{PpEqvEWconv2CUCompFt} \ {\rm(\cite{MTY09b})} \ 
Let $\{\mu_m\}$ and $\mu$ be computable 
probability distributions. Then the effective convergence of $\{\mu_m\}$ to $\mu$ 
is equivalent to the effective weak convergence of $\{\mu_m\}$ to $\mu$, 
that is,  $\{\mu_m(f_n)\}$ converges to $\mu(f_n)$  effectively 
for all effectively bounded computable sequence of functions. 
\end{prop}

For characteristic functions, the following classical Bochner's Theorem holds. 

\begin{thm}\label{ThCllBochner} \ 
{\rm(Bochner's Theorem \cite{Ito78})} \  
In order for $\varphi(t)$ to be a characteristic function, it is necessary 
and sufficient that the following three conditions hold. 

{\rm(i)} \ $\varphi$ is positive definite. \ 
{\rm(ii)} \ $\varphi(t)$ is continuous at 0. \ 
{\rm(iii)} \ $\varphi(0)=1.$
\end{thm}

\section{Effective Integrability}

Effective integrability of Fine computable functions on $[0,1)$ and $[0,1)^2$ 
with respect to the Lebesgue measure $dx$ and $dx\times dy$ respectively 
has been treated in \cite{MTY07b}, \cite{MTY08a}, \cite{MTY08b}, \cite{MTY09a}. 
We need to extend some of these results to 
$(\MR, dx)$, $(\MR, \mu)$ and $(\MR^2, \mu(dx)dy)$ for computable integrands, 
where $\mu(dx)dy$ is the 
product measure of a probability distribution $\mu$ and the Lebesgue measure 
$dx$. This product measure is defined by assigning the measure 
$\mu((a,b])\times(d-c)$ to the product set $(a,b]\times(c,d]$.

A bounded measurable function is integrable on any finite interval. 
A positive (non-negative) measurable function $f$ is called 
integrable if $\{\int_{[-n,n]}(f\wedge m)(x)dx\}$ converges to a finite limit 
as $m$ and $n$ tend to infinity, where $(f\wedge m)(x)=\min\{f(x), m\}$. 
We denote this limit with $\int_{\MR}f(x)dx$. 

A measurable $f$ is called integrable 
if $f^+$ and $f^-$ are integrable, where $f^+(x)=f(x)\vee 0$ and 
$f^-(x)=(-f(x))\vee 0$. 
$\int_{\MR}f(x)dx$ is then defined to be 
$\int_{\MR}f^+(x)dx-\int_{\MR}f^-(x)dx$. 

If $f$ is integrable, then $\int_{\MR}|f(x)|w^c_n(x)dx$ converges to zero. 

$\int_Ef(x)dx$ is defined to be $\int_{\MR}\chi_E(x)f(x)dx$. 

A continuous function on a finite closed interval is Riemann integrable, 
and it is also Lebesgue integrable. The two integrals coincide. 

About effective integrability of computable functions, we have the following 
theorem.

\begin{thm}\label{ThEIntBFCCI} \ {\em(\cite{PR88})} \ 
Let $\{f_n\}$ be a computable sequence of functions. 
\begin{enumerate}[\em(i)]
\item
$\{\int_{[a_m,b_m]}f_n(x)dx\}$ is a computable double sequence of reals 
for computable sequences $\{a_m\}$ and $\{b_m\}$. 

\item If $\{f_n\}$ converges effectively to $f$, 
then  $\{\int_{[a_m,b_m]}f_n(x)dx\}$ converges effectively to \\
$\{\int_{[a_m,b_m]}f(x)dx\}$ effectively in $m$. 
\end{enumerate}
\end{thm}

\begin{defi}\label{DefCompInt4Seq} \ 
({\em Effective integrability})
A sequence of computable functions $\{f_n\}$ is said 
to be effectively integrable (on $\MR$) if  the sequences 
$\{\int_{[-m,m]}f_n^+(x)dx\}$ and $\{\int_{[-m,m]}f_n^-(x)dx\}$ 
converge effectively as $m$ tends to infinity effectively in $n$. 

We denote these limits with $\int_{\MR}f_n^+(x)dx$ and $\int_{\MR}f_n^-(x)dx$. 
$\int_{\MR}f_n(x)dx$ is defined to be 
$\int_{\MR}f_n^+(x)dx-\int_{\MR}f_n^-(x)dx$. 
\end{defi}

We can easily obtain the following proposition.

\begin{prop}\label{PpEffIntBCF} \ 
A computable sequence of functions $\{f_n\}$ is effectively integrable 
if and only if $\int_{\MR}|f_n(x)|w^c_m(x)dx$ converges effectively 
to 0 as $m$ tends to infinity effectively in $n$. 
\end{prop}

By Proposition \ref{PpEffIntBCF} and Theorem \ref{ThEIntBFCCI}, 
we can prove the following theorem.

\begin{thm}\label{ThEDCT} \ 
{\rm(Effective dominated convergence theorem for $dx$)} \ 
Let $\{g_{m,n}\}$ be a computable sequence of functions which converges 
effectively to $\{f_m\}$. 
Assume that there exists an effectively integrable computable sequence of 
functions $\{h_m\}$ such that $|g_{m,n}(x)|\leq h_m(x)$.  

Then 
$\{g_{m,n}\}$ is effectively integrable and $\{\int_{\MR}g_{m,n}(x)dx\}$ 
converges effectively to \\ $\{\int_{\MR}f_m(x)dx\}$ as $n$ tends to infinity 
effectively in $m$.  
\end{thm}

For the proof of the theorems in the next section, we treat 
$\int_{\MR}f(x,y)dy$ and \\ $\int_{\MR}f(x,y)\mu(dy)$ for a computable 
binary function $f(x,y)$. 

We say that a sequence of binary functions $\{f_n\}$  is computable 
with compact support $\{L(n)\}$, if $\{f_n\}$ is (uniformly) computable 
and $L(n)$ is a recursive function satisfying 
$f_n(x,y)=0$ for all $x,y$ with $\min\{|x|, |y|\}\geq L(n)$.

\begin{thm}\label{ThEffKernel} \ 
Let $\{f_n(x,y)\}$ be a  computable sequence of binary functions 
and let $\{\mu_m\}$  be  a computable sequence 
of probability distributions.  

\begin{enumerate}[\em(1)]
\item If $\{f_n(x,y)\}$ is effectively bounded, then, 
as a function of $x$, $\{\int_{\MR}f_n(x,y)\mu_m(dy)\}$ is an effectively  
bounded computable double sequence of functions. 

\item If there exists an effectively integrable computable function 
$g(y)$such that $|f_n(x,y)|\leq g(y)$, then $\{\int_{\MR}f_n(x,y)dy\}$ is 
a computable sequence of functions on the real line. 
\end{enumerate}
\end{thm}

\noindent{\it Proof.} \ (1) \ {\it Sequential computability}: \ 
Let $\{x_{\ell}\}$ be a computable sequence 
of reals. Then, $\{f_n(x_{\ell},y)\}$ is a computable (double) sequence 
of functions of $y$. 
By the computability of $\mu_m$ and Proposition \ref{PpEqvWcomp2CUCompFt}, 
$\{\int_{\MR}f_n(x_{\ell},y)\mu_m(dy)\}$ is computable. 

{\it Effective continuity}: 
Let $\alpha(n,H,k)$ be a recursive modulus of continuity of $\{f_n\}$, 
$L(m,k)$ be a recursive function in Lemma \ref{LmEffTight1} 
and let $M(n)$ be a recursive bound of $\{f_n\}$. 
Then, $\mu_m(w^c_{L(m,k)})<2^{-k}$. 
If $|x|,|z|, |y|,|w|\leq H$ and $|x-y|,|z-w|<2^{-\alpha(n,H,k)}$, 
then $|f_n(x,z)-f_n(y,w)|<2^{-k}$.

Put $\ell=\ell(m,k,H)=\max\{H, L(m,k)\}$ and 
assume 
$|x-y|<2^{-\alpha(n,\ell,k)}$. Then, \\ 
$\mu_m(w^c_{\ell})\leq \mu_m(w^c_{L(m,k)})<2^{-k}$ and 
\begin{eqnarray*} 
&&\textstyle  |\int_{\MR}f_n(x,z)\mu_m(dz)-\int_{\MR}f_n(y,z)\mu_m(dz)|\\ 
&\leq&\textstyle  \int_{\MR}|f_n(x,z)-f_n(y,z)|w_{\ell}(z)\mu_m(dz)
+\int_{\MR}|f_n(x,z)-f_n(y,z)|w^c_{\ell}(z)\mu_m(dz) \\ 
&\leq&\textstyle  \int_{\MR}|f_n(x,z)-f_n(y,z)|w_{\ell}(z)\mu_m(dz)
+2M(n)2^{-k} \\ 
&\leq&\textstyle (1+2M(n))2^{-k}. 
\end{eqnarray*}
Hence $\{\int_{\MR}f_n(x,z)\mu_m(dz)\}$ is effectively continuous with respect to 
\[\gamma(m,n,H,k)=\alpha(n,\ell(m,k,H),k+2M(n)+3)\ .\] \smallskip
(2) can be proved similarly by means of Theorem \ref{ThEIntBFCCI} and 
Proposition \ref{PpEffIntBCF}. 
\qed

\vspace{0.5em}

We explain at this point the computability of a complex-valued function. 
$i$ denotes $\sqrt{-1}$, $\mathrm{Re(z)}$ denotes the real part of $z$ and 
$\mathrm{Im}(z)$ denotes the imaginary part of $z$. 

A sequence of complex numbers $\{r_n+is_n\}$ is called recursive  
if $\{r_n\}$ and $\{s_n\}$ are recursive sequences of rationals numbers. 
Computability of  a sequence of complex numbers $\{z_n\}$ is defined 
in terms of recursive sequences of complex numbers and it is equivalent to 
computability of $\{\mathrm{Re}(z_n)\}$ and $\{\mathrm{Im}(z_n)\}$.

Computability of a complex-valued function on $\MR$ is defined by 
viewing it as a mapping from the metric space $\MR$ to the metric space 
$\MC$ (\cite{MTY96}). 
Computability of a sequence of complex-valued functions $\{f_n\}$ is 
defined similarly and is equivalent to computability of 
$\{\mathrm{Re}(f_n)\}$ and $\{\mathrm{Im}(f_n)\}$. 
$\{f_n\}$ is uniformly computable if and only if $\{\mathrm{Re}(f_n)\}$ and 
$\{\mathrm{Im}(f_n)\}$ are uniformly computable. 

A  complex-valued function is called integrable if $\{\mathrm{Re}(f)\}$ and 
$\{\mathrm{Im}(f)\}$ are integrable and $\int_{\MR}f(x)dx$ is defined 
to be $\int_{\MR}\mathrm{Re}(f)(x)dx+i\int_{\MR}\mathrm{Im}(f)(x)dx$. 

A sequence of complex-valued functions $\{f_n\}$ is said to be 
effectively integrable if each $f_n$ is integrable and 
$\{\int_{\MR}\mathrm{Re}(f_n)(x)dx\}$ and $\{\int_{\MR}\mathrm{Im}(f_n)(x)dx\}$ 
are computable sequences of reals. 

Theorem \ref{ThEffKernel} can be generalized 
easily to complex-valued functions.

\section{Characteristic functions}
\label{SecCharFun}

\begin{thm}\label{ThCompDist2ComChFt}\ 
If a sequence of probability distributions $\{\mu_m\}$ is computable, 
then the corresponding sequence of characteristic functions 
$\{\varphi_m\}$ is uniformly computable. 
\end{thm}

\noindent
{\it Proof.} \ {\it Sequential computability:} \ Let $\{t_n\}$ be a
computable sequence of reals.  Then $\{e^{it_nx}\}$ is a bounded
computable sequence of functions of $x$. Hence the extension of
Proposition \ref{PpEqvWcomp2CUCompFt} to the complex case shows that
$\{\mu_n(e^{it_nx})\}$ is a computable sequence of complex numbers.

{\it Effective uniform continuity:} \ 
It is well known that 
$|\varphi_m(t+h)-\varphi_m(t)|\leq \mu_m(|e^{ihx}-1|)$ 
and $|e^{iz}-1|\leq|z|$ for any real $z$. 

Take a recursive function $L(m,k)$ 
in Lemma \ref{LmEffTight1}.  Then, we obtain 
\begin{eqnarray*}
 && \mu_m(|e^{ihx}-1|)
= \mu(w_{L(m,k)}|e^{ihx}-1|)+\mu(w^c_{L(m,k)}|e^{ihx}-1|) \\ 
&\leq& \textstyle  \int_{\MR}w_{L(m,k)}(x)|hx|\mu_m(dx) +2\cdot 2^{-k} 
\leq |h|(L(m,k)+1)\mu_m(w_{L(m,k)})+2\cdot 2^{-k} \\ 
&\leq& |h|(L(m,k)+1)+2\cdot 2^{-k}
\end{eqnarray*}
So, if $|h|<\frac{1}{(L(m,k+2)+1)2^{k+2}}$, then 
$|\varphi_m(t+h)-\varphi_m(t)|<2^{-k}$. 
\qed

\begin{thm}\label{ThEffGlivenko} \ 
{\rm (Effective Glivenko's theorem, cf. Theorem 2.6.4 in \cite{Ito78})}\ 
Let $\{\varphi_m\}$ and $\varphi$ be a computable sequence of 
characteristic functions and a characteristic function respectively, 
and let $\{\mu_m\}$ and $\mu$ be the corresponding probability distributions. 
Then,  $\{\mu_m\}$ converges effectively to $\mu$ if 
$\{\varphi_m\}$ converges effectively to $\varphi$. 
\end{thm}

\noindent
{\it Proof.} \ We follow the proof of Theorem 2.6.4 in \cite{Ito78} and 
prove that $\{\mu_m(f)\}$ converges effectively to $\mu(f)$ 
for any computable function $f$ with compact support. 
Notice that 
\begin{equation}\label{EqFubiniGnPhi}\textstyle 
\int_{\MR}\bigl(\int_{\MR}e^{izx}g(z)dz\bigr)\mu(dx)
=\int\hspace*{-0.3em}\int_{\MR\times\MR}g(z)e^{izx}\mu(dx)dz
=\int_{\MR}\varphi(z)g(z)dz
\end{equation}
holds for an integrable function $g$ and any pair of a probability 
distribution $\mu$ and the corresponding characteristic function 
$\varphi$ by virtue of the classical Fubini Theorem. 
Assume that $g$ is an effectively integrable computable function. Then 
$\int_{\MR}e^{izx}g(z)dz$ is a bounded uniformly computable function. 

Effective convergence of $\{\varphi_m\}$ to $\varphi$ implies
effective convergence of $\{\int_{\MR}\varphi_m(z)g(z)dz\}$ to
$\int_{\MR}\varphi(z)g(z)dz$ by Theorem \ref{ThEDCT}.
This proves that $\{\mu_m(h)\}$ converges effectively to $\mu(h)$ 
for functions of type $h(x)=\int_{\MR}e^{izx}g(z)dz$ 
with effectively integrable computable $g$. 

To complete the proof, it is sufficient to prove that 
any computable function $f$ with compact support can be 
approximated effectively by a sequence of functions $\{h_n\}$ 
of type of $h$. 

Now, define 
$ g_n(z)=\frac{1}{2\pi}e^{-\frac{z^2}{2n}}\int_{\MR}e^{-izy}f(y)dy$ 
and $h_n(x)=\int_{\MR}e^{izx}g_n(z)dz$. 
Then, $\{g_n\}$ is a computable sequence of functions,
and since
$|g_n(z)|\leq\frac{1}{2\pi}e^{-\frac{z^2}{2n}}\int_{\MR}|f(y)|dy$, it is effectively integrable. 
We have 
\begin{eqnarray*}
h_n(x)&=&\textstyle \int_{\MR}e^{izx}\frac{1}{2\pi}e^{-\frac{z^2}{2n}}dz
\int_{\MR}e^{-izy}f(y)dy
=\frac{1}{2\pi}\int_{\MR}f(y)dy\int_{\MR}e^{iz(x-y)}e^{-\frac{z^2}{2n}}dz\\ 
&=&\textstyle  \sqrt{\frac{n}{2\pi}}\int_{\MR}f(y)dy\int_{\MR}e^{iz(x-y)}
\frac{1}{\sqrt{2\pi n}}e^{-\frac{z^2}{2n}}dz
=\sqrt{\frac{n}{2\pi}}\int_{\MR}e^{-\frac{n(x-y)^2}{2}}f(y)dy\\ 
&=&\textstyle \frac{1}{\sqrt{2\pi}}\int_{\MR}
e^{-\frac{t^2}{2}}f(x+\frac{t}{\sqrt{n}})dt.  
\end{eqnarray*}
Hence, we obtain the following estimate: 
\begin{eqnarray*}\textstyle 
&& |h_n(x)-f(x)|  
= \textstyle \frac{1}{\sqrt{2\pi}}
\int_{\MR}e^{-\frac{t^2}{2}}|f(x+\frac{t}{\sqrt{n}})-f(x)|dt \\ 
&=&\textstyle  \frac{1}{\sqrt{2\pi}}\int_{|t|\leq L}e^{-\frac{t^2}{2}}
|f(x+\frac{t}{\sqrt{n}})-f(x)|dt
+ \frac{1}{\sqrt{2\pi}}\int_{|t|>L}e^{-\frac{t^2}{2}}|f(x+\frac{t}{\sqrt{n}})-f(x)|dt \\ 
&\leq&\textstyle  
\sup_{|t|\leq L}|f(x+\frac{t}{\sqrt{n}})-f(x)|
+\frac{2}{\sqrt{2\pi}}M_f \int_{|t|>L}e^{-\frac{t^2}{2}}dt,
\end{eqnarray*} 
for any $L$, where $M_f$ denotes the maximum of $f$, which is computable.  
Furthermore, 
\begin{equation*}\textstyle 
\int_{|t|>L}e^{-\frac{t^2}{2}}dt \leq 
e^{-\frac{L^2}{2}}\int_{\MR}e^{-\frac{t^2}{2}}dt
=\sqrt{2\pi}e^{-\frac{L^2}{2}}.    
\end{equation*} 

Finally, let $\alpha(k)$ be a recursive modulus of continuity of $f$. 
For each $k$, take $L=L(k)$ so large that 
$2M_f e^{-\frac{L^2}{2}}<\frac{1}{2^{k+1}}$, 
and take $n>L^22^{2\alpha(k+1)}$. Then, we have 

$|h_n(x)-f(x)|<\frac{1}{2^k}$. 
\qed

\begin{thm}\label{ThEffLevy} \ 
Assume that a sequence of probability distributions $\{\mu_m\}$ 
converges effectively to a probability distribution $\mu$. 
Then the corresponding sequence of characteristic functions 
$\{\varphi_m\}$ converges effectively to $\varphi$, which is the 
corresponding characteristic function of $\mu$. 
\end{thm}

If a sequence of probability distributions $\{\mu_m\}$ 
converges effectively to a probability distribution $\mu$, then 
we can replace $L(m,k)$ with $\alpha(k)$ in the proof of 
Theorem \ref{ThCompDist2ComChFt}. Hence, we obtain the following lemma. 

\begin{lem}\label{LmEffUnifEqCont} \ 
Assume that a sequence of probability distributions $\{\mu_m\}$ 
converges effectively to a probability distribution $\mu$. 
Then the corresponding characteristic functions 
$\{\varphi_m\}$ and $\varphi$ are effectively uniformly equi-continuous, 
that is, there exists a recursive function $\beta(k)$ such that 
$|t-s|<2^{-\beta(k)}$ implies $|\varphi_m(t)-\varphi_m(s)|<2^{-k}$ 
for any $m$ and $|\varphi(t)-\varphi(s)|<2^{-k}$. 
\end{lem}

\noindent 
{\it Proof of Theorem \ref{ThEffLevy}.} \ 
Let $\{e_{\ell}\}$ be an effective enumaration of all dyadic rationals and 
$\beta(k)$ be an effective modulus of uniform equi-continuity in 
Lemma \ref{LmEffUnifEqCont}. 

$\{e^{ie_{\ell}x}\}$ is an effectively bounded computable sequence of
functions. Hence, $\{\mu_m(e^{ie_{\ell}x})\}$ converges effectively to
$\mu(e^{ie_{\ell}x})$ by the extension of Proposition
\ref{PpEqvEWconv2CUCompFt} to the complex case, that is, there exists
a recursive function $\gamma(\ell,k)$ such that $m \geq
\gamma(\ell,k)$ implies
$|\mu_m(e^{ie_{\ell}x})-\mu(e^{ie_{\ell}x})|<2^{-k}$.

Let us assume $|t|\leq M$. Put $t_j=-M+j\,2^{-\beta(k)}$ 
(where $0\leq j \leq 2M2^{\beta(k)}$). 
Then $t \in [t_j,t_{j+1})$ for some $j$. 
We can find a recursive function $\xi(j)$ such that $t_j=e_{\xi(j)}$. 
Define 

\hspace*{1em}
$\eta(M,k)=\max\{\gamma(\xi(0),\beta(k)),\gamma(\xi(1),\beta(k)),
\ldots,\gamma(\xi(2M2^{\beta(k)}),\beta(k))\}$. 

Suppose that $|t| \leq M$ and $m \geq \eta(M,k)$. Then 

\hspace*{1em}
$|\varphi_m(t)-\varphi(t)| \leq |\varphi_m(t)-\varphi_m(t_j)|+
|\varphi_m(t_j)-\varphi(t_j)|+|\varphi(t_j)-\varphi(t)|
<3\cdot 2^{-k}.$ \\ 
Therefore $\{\varphi_m(t)\}$ converges effectively to $\varphi(t)$. 
\qed

\vspace{1em}

\noindent Modifying the proof of Bochner's Theorem in \cite{Ito78}, 
we can prove the following two theorems.

\begin{thm}\label{ThComPhi2Mu}\ 
Let $\{\mu_m\}$  be  probability distributions, and let 
$\{\varphi_m\}$  be the corresponding characteristic functions. 
Then, $\{\mu_m\}$ is computable if $\{\varphi_m\}$ is computable. 
\end{thm}

\noindent
{\it Proof.} \  We prove the theorem for a single $\mu$ and the 
corresponding characteristic function $\varphi$. 
First, we note that the following is proved in \cite{Ito78}: 
If we put 
\begin{equation*}\textstyle 
f_n(x)=\frac{1}{2\pi}\int_{\MR}\varphi(t)e^{-\frac{t^2}{n}}e^{-ixt}dt, 
\end{equation*}
then $f_n(x)\geq 0$ and $\int_{\MR}f_n(x)dx=1$. 
By computability of $\varphi$ and  Theorem \ref{ThEffKernel} (2), 
$\{f_n\}$ is a computable sequence of functions. 

Define $\nu_n(dx)=f_n(x)dx$. Then, $\{\nu_n\}$ is a computable sequence 
of probability distributions. 
$\psi_n(z)=\varphi(z)e^{-\frac{z^2}{n}}$ 
is the  characteristic function of $\nu_n$. 

Computability of $\{\psi_n\}$ and 
effective convergence of $\{\psi_n\}$ to $\varphi$ are obvious. 

Therefore, $\nu_n$ converges effectively to $\mu$ by 
Theorem \ref{ThEffGlivenko} and hence $\mu$ is computable 
by Proposition \ref{PpLimEffComvDist}. 

The argument above goes through for a sequence $\{\mu_m\}$. 
\qed 

\vspace{0.5em}

\noindent Combining Theorems \ref{ThCompDist2ComChFt} and \ref{ThComPhi2Mu}, 
we obtain the following theorem.

\begin{thm}\label{ThEquivCom}\ 
Let $\{\mu_m\}$ be  probability distributions, 
and let $\{\varphi_m\}$ be the corresponding characteristic 
functions.

Then, $\{\mu_m\}$ is computable if and only if $\{\varphi_m\}$ is computable. 
\end{thm}

\begin{thm}\label{ThEffBochner}\ {\rm(Effective Bochner's theorem)} \ 
In order for $\varphi(t)$ to be the characteristic function of a 
computable probability distribution, it is necessary 
and sufficient that the following three conditions hold. 

{\rm(i)} \ $\varphi$ is positive definite. \ 
{\rm(ii${}'$)} \ $\varphi$ is computable. \ 
{\rm(iii)} \ $\varphi(0)=1.$
\end{thm}

\noindent 
{\it Proof.} \ $\varphi$ is a characteristic function of a 
probability distribution if and only if the conditions (i), (ii), and (iii) 
of Theorem \ref{ThCllBochner} hold. 
If (ii) is replaced by (ii${}'$), the corresponding probability distribution 
is computable by Theorem \ref{ThComPhi2Mu}. 

The converse is a special case of Theorem \ref{ThCompDist2ComChFt}. 
\qed

% {\bf pause}

\begin{example} \rm 
For $\delta_0$ and $\mu_m(dx)=2^{m-1}\chi_{[-2^{-m},2^{-m}]}dx$, 
$\{\mu_m\}$ converges effectively to $\delta_0$. 

% The corresponding probability distribution functions are Fine computable 
% (cf. \cite{MTY11}). 

$\varphi_{\delta_0}(t)\equiv 1$. 

$\varphi_m(t)=2^{m-1}\int_{-2^{-m}}^{2^{-m}}e^{itx}dx
=2^m\int_0^{2^{-m}}\cos tx dx=\frac{\sin t2^{-m}}{t2^{-m}}$ 
if $t\neq 0$, and $\varphi_m(0)=1$. 

$\{\varphi_m(t)\}$ converges effectively to 1. 

\end{example}

\section{De Moivre-Laplace Central Limit Theorem}

Let $(\Omega,\CB,\MP,\{X_n\})$ be a realization of Coin Tossing 
with success probability $p$, that is, $(\Omega,\CB,\MP)$ is a 
probability space and $\{X_n\}$ is a sequence of independent 
$\{0, 1\}$-valued random variables with the same probability distribution  
$\MP(X_n=1)=p$ and $\MP(X_n=0)=q=1-p$. 

The probability distribution of $S_m=X_1+\cdots+X_m$ is 
the binomial distribution 
$\mu_m=\sum_{\ell=0}^m \binom{m}{\ell}p^{\ell}(1-p)^{m-\ell}\delta_{\ell}$  
and its characteristic function 
$\varphi_m(t)$ is equal to $(pe^{it}+q)^m$.  

\begin{thm}\label{ThDMLCLT} \ {\rm(Effective de Moivre-Laplace 
central limit theorem)} \ If $p$ is a computable real number, then the
  sequence of probability distributions of $Y_m=\frac{X_1+\cdots X_m
    -mp}{\sqrt{mp(1-p)}}
  =\sum_{\ell=1}^m\frac{X_{\ell}-p}{\sqrt{mpq}}$ converges effectively
  to the standard Gaussian distribution.
\end{thm}

\begin{lem}\label{LmEstExp} \ {\rm(Ito \cite{Ito78}) }\hfill
\begin{enumerate}[\em(1)] 
\item Put $R_n(t)=\int_0^t\int_0^{t_1}\cdots\int_0^{t_{n-1}}i^{n}e^{it_{n}}
dt_n\cdots dt_1$. Then \\ 
$e^{it}=\sum_{k=0}^{n-1}\frac{(it)^k}{k!}+R_n(t)$  and 
$|R_n(t)|\leq \frac{|t|^{n}}{(n)!}$ for any $t\in\MR$. 

\item  If $|z|\leq\frac{1}{2}$, then $|\log (1+z)-z|\leq |z|^2$. 
\end{enumerate}
\end{lem}

{\it Outline of the proof of Theorem \ref{ThDMLCLT}:} \ By virtue of
Theorem \ref{ThEffGlivenko}, it is sufficient to prove effective
convergence of $\log \varphi_m(t)$ to $-\frac{t^2}{2}$, since then the
corresponding districution is the standard Gaussian.  We follow Ito
\cite{Ito78} (p. 192).  The following holds by definition and Lemma
\ref{LmEstExp} (1);
\begin{eqnarray*}
\psi_m(t) &=& \ME(e^{itY_m})=\textstyle 
\prod_{\ell=1}^m\ME(e^{it\,\frac{X_{\ell}-p}{\sqrt{mpq}}})
% =(pe^{\frac{itq}{\sqrt{mpq}}}+qe^{-\,\frac{itp}{\sqrt{mpq}}})^m
=(pe^{\frac{it\sqrt{q}}{\sqrt{mp}}}+qe^{-\,\frac{it\sqrt{p}}{\sqrt{mq}}})^m. \\ 
e^{\frac{it\sqrt{q}}{\sqrt{mp}}} &=&\textstyle 1+it\frac{\sqrt{q}}{\sqrt{mp}}
-t^2\frac{q}{2mp}+R_3(\frac{it\sqrt{q}}{\sqrt{mp}}). \\ 
e^{-\,\frac{it\sqrt{p}}{\sqrt{mq}}} &=&\textstyle  1-it\frac{\sqrt{p}}{\sqrt{mq}}
-t^2\frac{p}{2mq}+R_3(-\,\frac{it\sqrt{p}}{\sqrt{mq}}). \\  
\log \psi_m(t) &=& \textstyle 
m\,\log\bigl(1-\frac{t^2}{2m}+R_3(\frac{it\sqrt{q}}{\sqrt{mp}})
+R_3(-\,\frac{it\sqrt{p}}{\sqrt{mq}})\bigr).\\ 
\textstyle |R_3(\frac{it\sqrt{q}}{\sqrt{mp}})|
&\leq& \textstyle   |t|^3\bigl(\frac{q}{mp}\bigr)^{\frac{3}{2}}. \\ 
\textstyle |R_3(-\,\frac{it\sqrt{p}}{\sqrt{mq}})| &\leq& \textstyle 
|t|^3\bigl(\frac{p}{mq}\bigr)^{\frac{3}{2}}. 
\end{eqnarray*}

\noindent 
If $\frac{t^2}{2m}+|t|^3\bigl(\frac{q}{mp}\bigr)^{\frac{3}{2}}+
|t|^3\bigl(\frac{p}{mq}\bigr)^{\frac{3}{2}}<\frac{1}{2}$, then 
by Lemma \ref{LmEstExp} (2) and the above facts, 

$|\log \psi_m(t)-(-\frac{t^2}{2})| \leq 
|t|^3\bigl\{\bigl(\frac{q}{p}\bigr)^{\frac{3}{2}}+
\bigl(\frac{p}{q}\bigr)^{\frac{3}{2}}\bigr\}\frac{1}{\sqrt{m}}+
\bigr(\frac{t^2}{2m}+|t|^3\bigl(\frac{q}{mp}\bigr)^{\frac{3}{2}}+
|t|^3\bigl(\frac{p}{mq}\bigr)^{\frac{3}{2}}\bigl)^2$. 

\noindent The last term converges to zero effectively and uniformly in
$|t|\leq K$ as $m$ tends to infinity.  \qed

\vspace{1em}

\noindent 
{\bf Acknowledgment: } The authors would like to express their gratitude 
to the referees for their valuable suggestions. 
Improvement of  Theorem \ref{ThEffLevy} and its proof especially owe to 
one of the referees.

\end{document}